\documentclass[useAMS,usenatbib]{mn2e}
\usepackage{aasmacros}
\usepackage{graphicx}
\usepackage{amsmath}
\usepackage{color}

\bibpunct{(}{)}{;}{a}{,}{,}

\let\cite=\citen
\def\apj{ApJ}
\def\apjs{ApJl}
\def\mnras{MNRAS}
\def\aaps{Astron. Astrophys. Suppl. Ser.}
\def\aap{Astronomy \& Astrophysics}
\def\aj{A.J.}
\def\aaps{Astron. Astrophys. Suppl. Ser.}
\def\aplett{Ap.J. (Letters)}
\def\prd{Phys. Rev.D}
\def\araa{Ann.Rev.Astr.Ap.} 
\def\nar{New Astronomy Reviews}
\def\aapr{Astronomy \& Astrophysics Reviews}
\def\apjs{Ap.J. (Suppl)}

\title[Contribution to the Diffuse Radio Background from Extragalactic Radio Sources ]{Contribution to the Diffuse Radio Background from Extragalactic Radio Sources}

\author[Vernstrom et
al.]{T. Vernstrom\thanks{E-mail:tvern@phas.ubc.ca}, Douglas Scott, 
  J.V. Wall  \\
  Department of Physics and Astronomy, University of British
  Columbia, Vancouver, BC Canada \\
}
\begin{document}
  
\date{2 February 2011}

\pagerange{\pageref{firstpage}--\pageref{lastpage}} \pubyear{2010}

\maketitle

\label{firstpage}

\begin{abstract} \

We examine the brightness of the Cosmic Radio Background (CRB)
by comparing the contribution from individual source counts to absolute
measurements. We use a
compilation of radio counts to estimate the contribution of detected
sources to the CRB in several different frequency bands.  Using a
Monte Carlo Markov Chain technique, we estimate the brightness values
and uncertainties, paying attention to various sources of systematic
error.  At $\nu$
= $150\,$MHz, $325\,$MHz, $408\,$MHz, $610\,$MHz, $1.4\,$GHz, $4.8\,$GHz, and
$8.4\,$GHz our calculated contributions to the background sky
temperature are 18, 2.8, 1.6, 0.71, 0.11, 0.0032, 0.0059 K, respectively. We then 
compare our results to absolute measurements from the
ARCADE 2 experiment.   
If the ARCADE 2 measurements are correct and come from
sources, then there must be an additional population of
radio galaxies, fainter than where current data are probing.  More
specifically, the Euclidean-normalized counts at 1.4 GHz have to have
an additional bump below about 10 $\mu$Jy.

\end{abstract}


\section{Introduction}
\label{sec:introduction}

Investigating what sources make up the diffuse extragalactic
background over a wide range of wavelengths can help us to understand
the different physical mechanisms which govern the generation and
transport of energy over cosmic
time (e.g \citealt{Longair69,Ressell90}). Much effort has
gone into resolving the sources which comprise the background at
$\gamma$-ray, X-ray, optical, and infra-red
wavelengths
(e.g. \citealp{Madau00,Hauser01,Brandt05,Lagache05}). 
However, the radio part of the spectrum has
received far less attention. While there have been many radio surveys
and compilations of source counts done over the years, 
there have been only a few attempts
at using these to obtain estimates of the background temperature
(\citealp{Longair66,Pooley68,Wall90,Burigana04}). 
With the advent of new absolute measurements of the radio
background, coupled with radio source counts to ever increasing depths,
the topic has undergone something of a revival.

Recently a paper by \citet{Gervasi08} attempted to obtain fits to
the source count data across a range of frequencies from $\nu$ = $150\,$
to $8440\,$MHz. From their fits, which ranged from $1\,\mu$Jy to $100\,$Jy,
they were able to integrate the source
counts to obtain an estimate of the sky brightness temperature
contribution at each of the frequencies. They determined a power-law 
sky brightness temperature dependency on frequency with a spectral
index of --2.7, which is in agreement with the frequency
dependence of the flux emitted by synchrotron dominated steep-spectrum
radio sources. These estimates were used to interpret absolute
measurements of the radio sky brightness by the TRIS experiment \citep{Zannoni08}.

More recently the results of the 2006 ARCADE 2 balloon-borne experiment
were released (\citealp{Seiffert09,Fixsen09}). This instrument
provided absolute measurements of the sky temperature at 3, 8, 10, 30,
and $90\,$GHz. These results showed a measured temperature of the radio
background about 5 times greater than that currently determined from
radio source counts, with the most notable excess of emission being
detected at $3\,$GHz. Since most systematic effects explaining this
emission were ruled out, we are left with the question of whether it
could be caused by some previously unknown source of extragalactic emission.  

It was suggested by Seiffert et al. in the ARCADE 2 
results paper that this excess emission may be coming from the sub-$\mu$Jy range.
One might imagine an unknown population of discrete sources existing
below the flux limit of current surveys. This issue was
further examined by \citet{Singal10}. Taking into account that a class
of low flux sources must extend to $\sim$ $10^{-2}$$\mu$Jy (at
$1.4\,$GHz), they concluded that this emission could primarily be coming from
ordinary star-forming galaxies at $z$ $>$ 1 if the radio to far-infrared
observed flux ratio increases with redshift. 

Before looking for radical causes of this emission,  it is worth
reexamining the observed radio source data to see if the ARCADE 2
result really does differ from what is expected. To do this we derive
new estimates of the source-integrated CRB at 
various frequencies and derive formal error estimates for each.
In Section \ref{sec:method} we describe the source count data
used, together with our procedure and results for fitting the observed radio
data. In Section \ref{sec:temp} we present our estimates for the background
sky temperature contributions and the analysis of the uncertainties associated
with these estimates. In Section \ref{sec:discussion} we compare our
results to those obtained by the ARCADE 2 and TRIS collaborations.

\section{The Radio Sources and Their Counts}
\label{sec:method}

\begin{figure*}

\includegraphics[scale=0.35, angle=270]{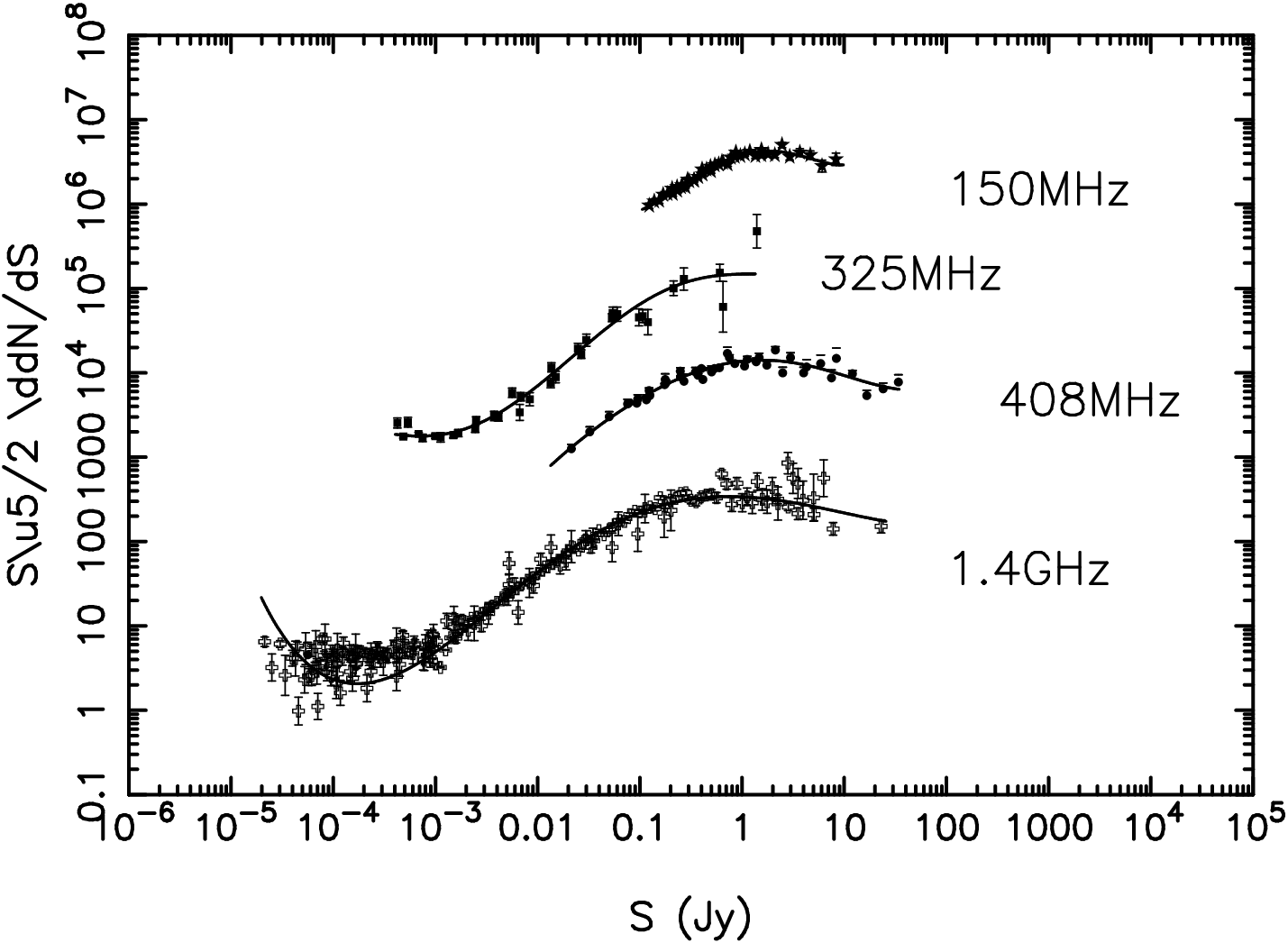}\includegraphics[scale=0.35, angle=270]{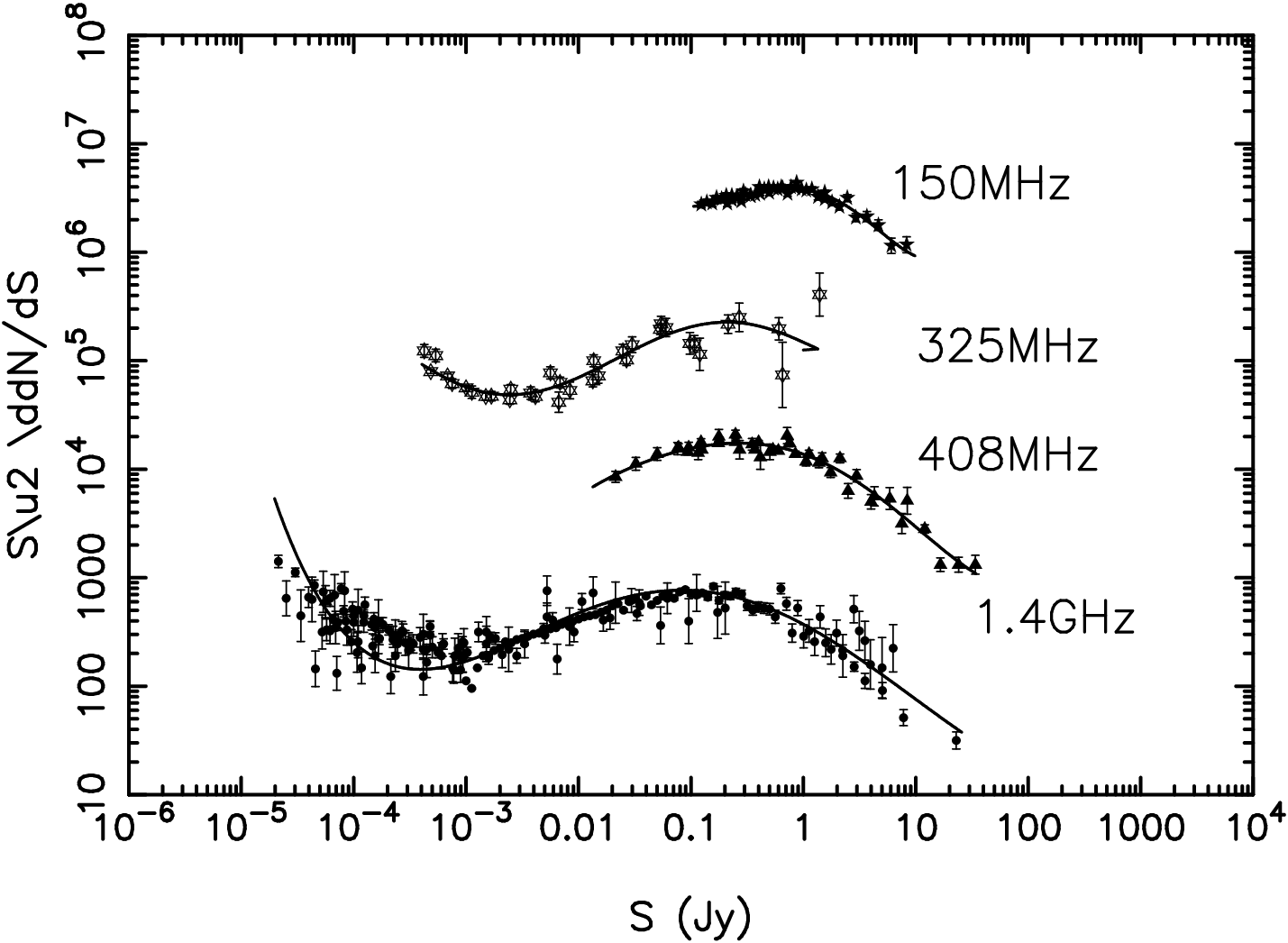}
\includegraphics[scale=0.35,angle=270]{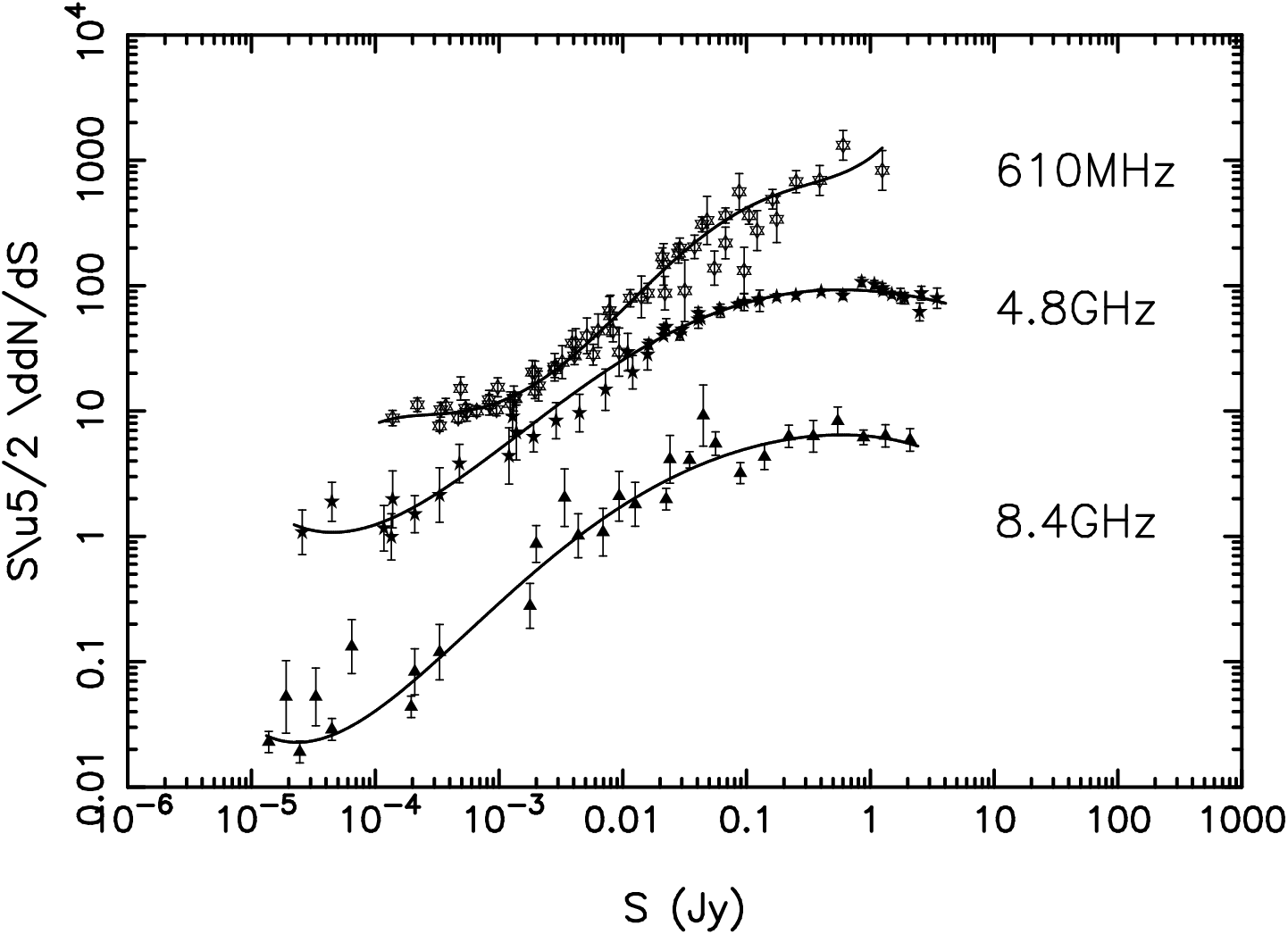}\includegraphics[scale=0.35,angle=270]{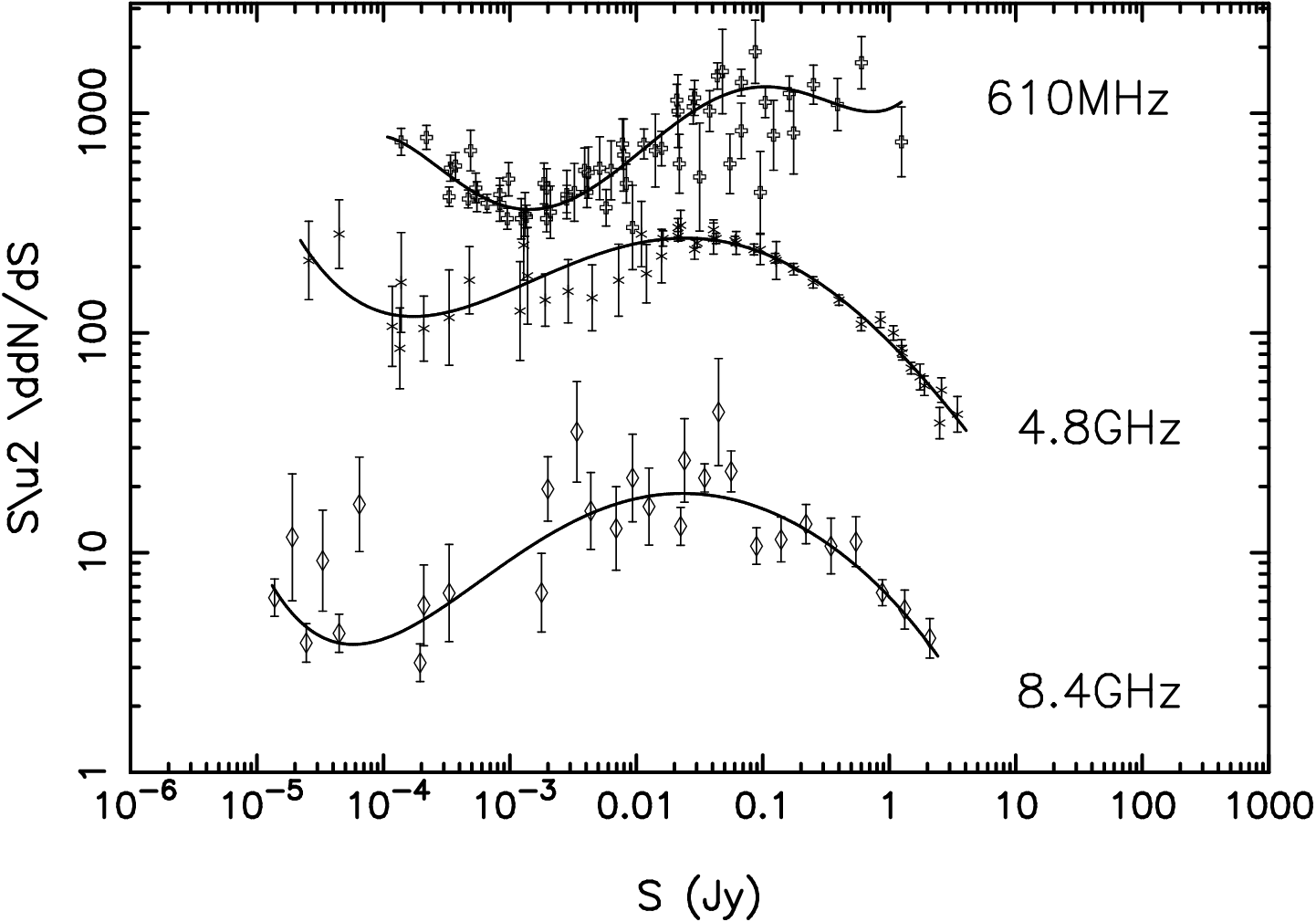}

\caption{Left: Differential source counts Euclidean normalized and
multiplied by c, with $c$ = 1000,100, 10, 1, 1, 1, and 0.1 
for $\nu$ = 150, 325, 408, 610, 1400, 4800, and $8400\,$MHz
respectively. Right: $S\textsuperscript{2}$ normalization to show 
where the contribution to the sky brightness temperature is
largest.The counts are a compilation from many different surveys,
listed in Table \ref{tbl:references}. Solid lines are best fit polynomials.}
\label{fig:best_fit}
\end{figure*}

\subsection{The Data Set}
\label{sec:data}

Radio source counts at lower frequencies
have been available since the 1960s. There are many compilations 
of radio source counts available, particularly in the last decade
(e.g. \citealp{Fomalont02, Bondi03, Hopkins03, Prandoni06}).
More recently deep continuum
surveys at higher frequencies have become available, and 
with the use of newer technologies, have dramatically increased the
amount and quality of data. The data used in this paper are
from continuum surveys carried out from 1979 to 2009 (see De
Zotti et al. 2009, Sirothia et al. 2009). We used source count
distributions from 150 MHz to 8400 MHz, with the individual
frequencies covered being $\nu$ = $150\,$MHz, $325\,$MHz, $408\,$MHz, 
$610\,$MHz, $1.4\,$GHz, $4.8\,$GHz and $8.4\,$GHz. 
References for all number counts used can be found in 
Table~\ref{tbl:references}.

\begin{table*}

\caption{References for the Extragalactic Radio Count Data Compilation}
\centering

\begin{tabular}{lll}\\

\hline

  Frequency &     &          References \\  
 
\hline

 $150\,$MHz & & \citealt{Hales88,McGil90}. \\
 $325\,$MHz & & \citealt{Owen08, Oort88, Sirothia09}.  \\
 $408\,$MHz & & \citealt{Benn82, Grueff88, Robertson73}. \\
 $610\,$MHz & & \citealt{Bondi07, Garn08, Ibar09}. \\ 
 & & \citealt{Katgert79, Moss07}. \\
 $1.4\,$GHz & & \citealt{Bondi08, Bridle72, Ciliegi99};
 \citealt{Fomalont06, Gruppioni99}; \\ & & \citealt{Hopkins03, Ibar09,
 Kellermann08, Mitchell85, Owen08};\\ 
 & & \citealt{Richards00, Seymour08, White97}. \\
 $4.8\,$GHz & & \citealt{Altschuler86, Donnelly87, Fomalont84,
 Gregory96} ;\\
 & & \citealt{Kuehr81, Pauliny80, Wrobel90}. \\
 $8.4\,$GHz & & \citealt{Fomalont02, Henkel05, Windhorst93}. \\

\hline
\end{tabular}\\
 \label{tbl:references}

\end{table*}

\subsection{The Number Counts Fit}
\label{sec:fit}

For fitting the source count data we opted to use a
fifth order polynomial. A third order polynomial was
used in source count fitting by \citet{Katgert88} and a sixth order
polynomial fit to the 1.4 GHz data was used by \citet{Hopkins03},
while \citet{Gervasi08} used simple power-law fitting.
Polynomial fits are simpler than some other choices of
function, but still allow for fitting of different
features in the data, such as the upturn at the low flux end seen at
some of the frequencies $($where we note that an additional sub-mJy
peak could make a substantial contribution to the background$)$. We
chose a fifth order polynomial as it is high enough order to account
for the features seen in the $1.4\,$GHz data. Going to higher
orders creates unneccesary extra parameters while not improving the
$\chi^2$ by a significant amount. 
Our empirical fits are performed on the Euclidean-normalized 
counts, i.e. $F(S)=S\textsuperscript{2.5}(dN/dS)$, with $S$ being the
flux density in Jy, using the polynomial with parameters
\begin{equation}
F(S)=A_0+A_1S+A_2S^2+A_3S^3+A_4S^4+A_5S^5 .
\end{equation}

The fitting is initially performed  using a $\chi^2$ minimization
routine. The $\chi^2$ minima are then 
used as starting points in a Monte Carlo Markov Chain, or
MCMC approach \citep{Lewis02}, which is used to refine the fits and obtain
estimates of uncertainty. More details 
on the MCMC method can be found in section~\ref{sec:mcmc}.
The best fit values for all the parameters at each of the 
frequency bands can be found in Table~\ref{tbl:chi} along with 
$\chi^2$ values for each fit. The data
and the best fit lines are plotted in Fig. ~\ref{fig:best_fit}, which
shows the Euclidean normalized data, as well as the
$S^2$ normalized results. These
$S^2(dN/dS)$ (surface brightness per logarithmic
interval in flux density) plots are included to show where the peak contributions to 
the background arises. The right-hand panels in Fig.~\ref{fig:best_fit}
show that the bulk of the background comes from relatively bright
radio sources, with $S$~$\sim$~1 Jy at the lowest frequencies to tens
of mJy at the highest frequencies. But there is a significant, and
still poorly characterized, contribution from much fainter sources.

Table \ref{tbl:chi} shows that
the $\chi^2$ values of the fits are generally
good, with all but one of the reduced $\chi^2$ values
being below 2. The exception is for the $1.4\,$GHz data set, with a 
$\chi^2$ of over 20 per degree of freedom.
To obtain anything like a reasonable $\chi^2$ 
we would have to increase the errors by  a factor of four. 
It is worrisome that the $1.4\,$GHz compilation is the one with the 
most available data.  
As can be seen in the plot, there are many data points that are
inconsistent with each other, even with the relatively large error
bars.

There are clearly systematic differences between different surveys at $1.4\,$GHz,
particularly at the faint end. In the $\mu$Jy range it is difficult to obtain
reliable counts, as this range is close to
the natural confusion limit of most radio surveys
(\citealp{Condon84,Windhorst85}) and hence the level of incompleteness
may be incorrectly estimated in some surveys. Moreover, at the bright
end there are significant and
systematic sources of error introduced when attempting to correct for
source extension and surface brightness limitations (see discussion
in \citealp{Singal10}). In addition to these effects, sampling variance
(enhanced by source clustering) can lead to differences in counts for
small fields. All of these systematic effects make it difficult to
assess robustly the uncertainties in the derived CRB, as we discuss in
the next section.

\begin{table*}

\caption{ $\chi^2$ values for best fits 
 at each of the frequencies}

\begin{tabular}{cccr@{.}lr@{.}lr@{.}lr@{.}lr@{.}lr@{.}l}\\

\hline
        $\nu$ & $\chi^2$ & Degrees of & \multicolumn{2}{c}{$A_0$}  & \multicolumn{2}{c}{$A_1$}  & \multicolumn{2}{c}{$A_2$}  & \multicolumn{2}{c}{$A_3$}  & \multicolumn{2}{c}{$A_4$}  &\multicolumn{2}{c}{$A_5$}  \\
        MHz &  & Freedom\\
\hline
        150 & 68 & 45 & 6&58 & 0&36 & $-0$&$65$ & $-0$&$19$ & 0&26 & 0&099\\        
        325 & 59 & 34 & 5&17 & 0&029 & $-0$&$11$ & 0&36 & 0&17 & 0&20\\
        408 & 66 & 44 & 4&13 & 0&13 & $-0$&$34$ & $-0$&$003$ & 0&035 & 0&01\\
        610 & 75 & 59 & 3&02  & 0&71 & 0&97 & 0&91 & 0&28 & 0&028\\
        1400 & 4230 & 196 & 2&53 & $-0$&$052$ & $-0$&$020$ & 0&051 & 0&010 & $-0$&$0013$\\
        4800 & 32 & 47 & 1&95 & $-0$&$076$ & $-0$&$15$ & 0&020 & 0&0029 & $-0$&$00079$\\
        8400 & 41 & 29 & 0&79 & $-0$&$10$ & $-0$&$23$ & $-0$&$051$ & $-0$&$019$ & $-0$&$0029$\\
\hline
\end{tabular}
\label{tbl:chi}

\end{table*}

\section{Contribution to Sky Brightness Temperature}

\label{sec:temp}

\subsection{Integration of Radio Counts}
\label{sec:integration}

We integrate best-fit polynomials to obtain the
contribution from the sources to the sky brightness. To do this we 
integrate the function $S(dN/dS)$ for each data set only in the range 
where data are available. We make this conservative choice
to avoid extrapolating at the very low and high flux density
ends. Because of this our estimates of the sky brightness should be seen as lower
limits. Thus to estimate the intensity we integrate
\begin{equation}
I(\nu)=\int^{S_{\rm{max}}}_{S_{\rm{min}}} \frac{dN}{dS}(\nu) \cdot S\ dS,
\end{equation}
where $S_{\rm{min}}$ and $S_{\rm{max}}$ are different 
for each frequency. Once the intensity is determined we use the 
Rayleigh-Jeans approximation to convert it to a brightness temperature,
\begin{equation}
T(\nu)=I(\nu) \frac{\lambda^2}{2k},
\end{equation}
where $k$ is the Boltzmann constant. The results from the
integration at each of the seven frequencies are listed in Table~\ref{tbl:temp}. 

\begin{table}
\centering
\caption{Values of the integrated sky brightness and temperature
contribution from radio source counts for different frequency bands. The
uncertainties are 1$\sigma$ limits determined from Markov chain
polynomial fits to the data. The high and low extrapolations are
discussed in the text.}
\begin{tabular}{llllll}\\
\hline
        $\nu$ &$\nu$$I_{\nu}$ & $T$  & $\delta T$
         & \multicolumn{2}{c}{Extrapolated $T$}   \\        
        & & & &  High & Low \\
        MHz & W $m^{-2}$ $\rm{sr}^{-1}$  & mK & mK & mK &mK\\
\hline
        150     & 1.8 $\times 10^{-14}$  & 17800       & 300    & 29400 & 18100\\
        325     & 2.1 $\times 10^{-14}$  & 2800        & 600    & 5040 & 3100\\
        408     & 2.9 $\times 10^{-14}$  & 1600        & 30     & 3000 & 1850\\ 
        610     & 4.2 $\times 10^{-14}$  & 710         & 90     & 1200 & 740\\
        1400    & 7.5 $\times 10^{-14}$  & 110         & 20     & 180 &  110\\
        4800    & 8.0 $\times 10^{-14}$  & 3.2         & 0.2     & 10.8 & 6.7\\
        8400    & 9.6 $\times 10^{-14}$  &0.59         & 0.05    & 3.0 & 1.9\\
\hline
\end{tabular}
\label{tbl:temp}
\end{table}

After obtaining these conservative estimates we next investigate the effect
of reasonable extrapolations on the results, with the limits of
integration broadened to $10^{-6}$ Jy and $10^2$ Jy for $S_{\rm{min}}$
and $S_{\rm{max}}$, respectively. The
$1.4\,$GHz data set has the most extensive coverage across the flux
density range. For this reason, we
manually extrapolate the curve for the $1.4\,$GHz data set and
integrate to get a new estimate for the background temperature. 
To extend the limits of integration for the $1.4\,$GHz data the end
behaviour of the polynomial is constrained with the assumption that
the counts fall off beyond the low-intensity end of the data. Artificial points
are added in this region and their positions varied until a
reasonable fit to the data is achieved. This procedure is repeated
with both a steep and shallow roll-off, to obtain high and low
background estimates. These slopes are chosen to be the most
reasonable steep and shallow estimates, with the
$\chi^2$s being a factor of 5 and 7 greater than the
best fit to the data alone.  The best fits for the extrapolations can be
seen in Fig. \ref{fig:extrap}. The higher estimate could have been
allowed to have an even shallower slope, therefore allowing for an even
higher background estimate; however anything much shallower than the chosen
fit would have $\chi^2$ values several times larger again.
This fact makes any shallower fits an unreasonable choice. The steep slope
estimate for the $1.4\,$GHz data ends up giving nearly the same result
for the background temperature as the unextrapolated estimate. This is
because the  unextrapolated estimate has a rising low-intensity tail
(Fig. 1), while the extrapolated estimate (Fig. 3) has limits of
integration extended with the roll-off procedure, whose end behaviour
is controlled to produce a steep downturn beyond the available data.

\begin{figure}

\includegraphics[scale=0.35,angle=270.0]{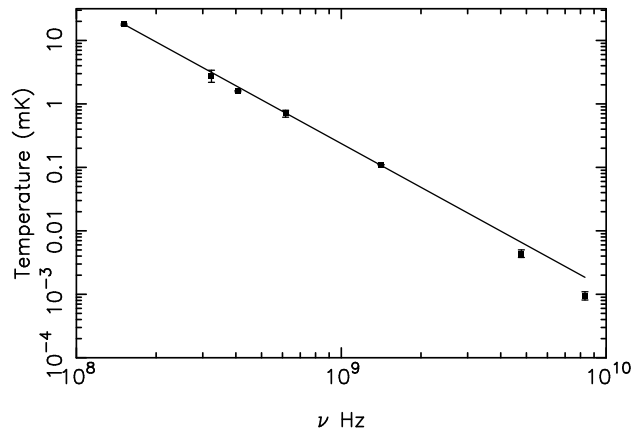}
\caption{Integration results and best fit power-law from
Equation \ref{eq:power}.}
\label{fig:power}
\end{figure}

\begin{figure}

\includegraphics[scale=0.35,angle=270.0]{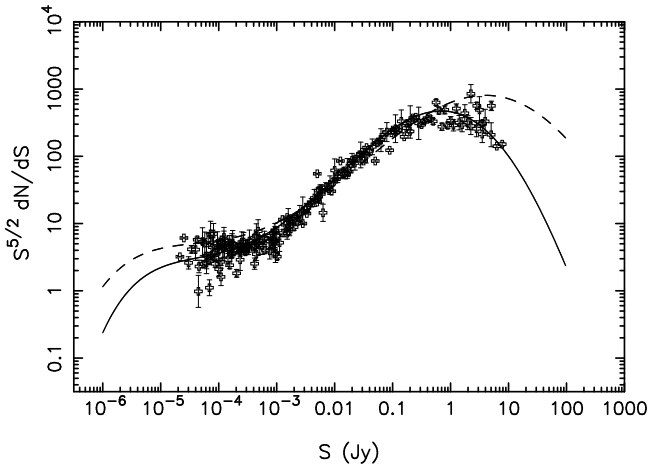}
\caption{  $1.4\,$GHz data set with fit lines showing 
           extrapolations out to $10^{-6}$
           and $10^2$Jy. The solid and dashed lines show
           estimates for steeper and shallower slopes, respectively. }
\label{fig:extrap}
\end{figure}

From our conservative estimate of the $1.4\,$GHz background a 
power-law is fit to the temperatures. This takes the form of
\begin{equation}
T(\nu)=A\left(\frac{\nu}{\rm1.4\, GHz}\right)^{\beta}\ ,
\label{eq:power}
\end{equation}
where $A$ is the power-law amplitude, and $\beta$ the
index. We set $A$ to the $1.4\,$GHz value of 0.110 K, while Monte
Carlo Markov Chains (section \ref{sec:mcmc}) were used to find the
best value of $\beta$ = --2.28 $\pm$ 0.1. The results of this power-law fit can
be seen in Fig. \ref{fig:power}. The fit is high for the two highest
frequency points, primarily as a result of the  $1.4\,$GHz
data, which has the most
flux density coverage and most available data.  It should be noted that this is just a
phenomenological fit; we see little merit in adopting a more complex
model such as a broken power-law just to satisfy two extreme data
points, particularly when this power-law is only used to obtain
extrapolation estimates.  

With the high and low extrapolation estimates
from the $1.4\,$GHz data, we use Equation \ref{eq:power} to obtain
estimates for the other frequencies. 
The results of the extrapolated estimates are given in
Table \ref{tbl:temp}. As even these reasonable extrapolations can 
change the background estimates by about a factor of two, it is clearly
important to push counts at all frequencies to fainter levels. 

\subsection{Uncertainty -- Monte Carlo Markov Chains}
\label{sec:mcmc}

\begin{figure}

\includegraphics[scale=0.35,angle=270.0]{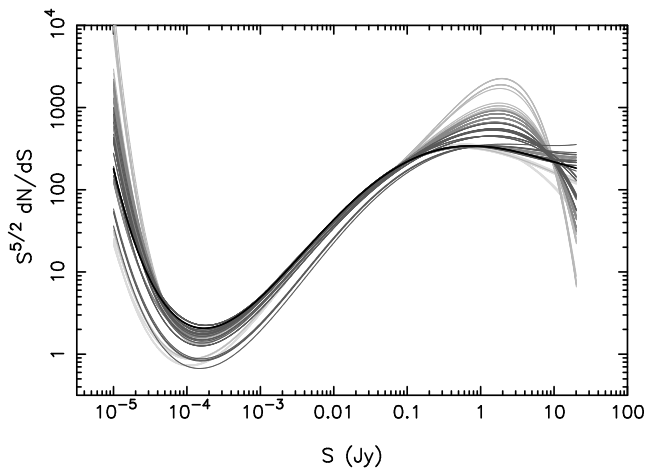}
\caption{  100 Markov Chain polynomial fits generated for the $1.4\,$GHz
           data set. Greyscale indicates relative probability, with the
           solid black line being the best fit curve.}
\label{fig:chains}
\end{figure}

To investigate the uncertainties thoroughly, we carry out our fits
with Monte Carlo Markov chains
for each of the data sets, using CosmoMC \citep{Lewis02} 
as a generic MCMC sampler. The $\chi^2$ function is sampled 
for each set using the polynomial in equation 1,
which is then fed to the sampler to locate the $\chi^2$ minimum. 
Each of the six parameters of the polynomials are varied for each step of
the chain and the chains are run with 500,000 steps. CosmoMC 
generates statistics for the chains, including the minimum $\chi^2$, 
the best fit values for each of the parameters, and their
uncertainties. As an example, Fig. \ref{fig:chains} shows different
polynomial fits tested by the MCMC and their relative probability for
the 1.4 GHz data set.

\begin{figure*}
\rotatebox{270}{
\includegraphics[width=55mm, height=65mm]{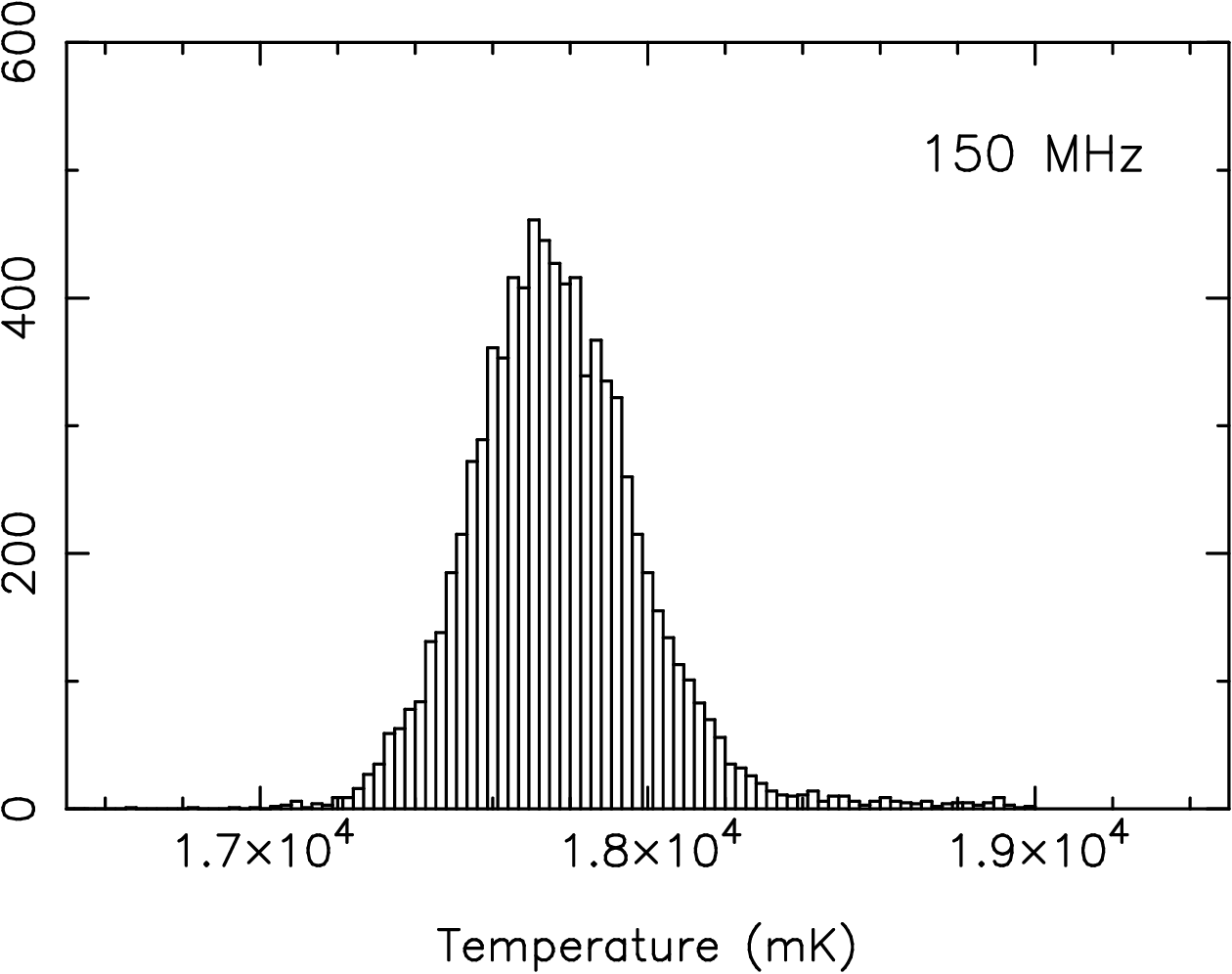}\includegraphics[width=55mm, height=65mm]{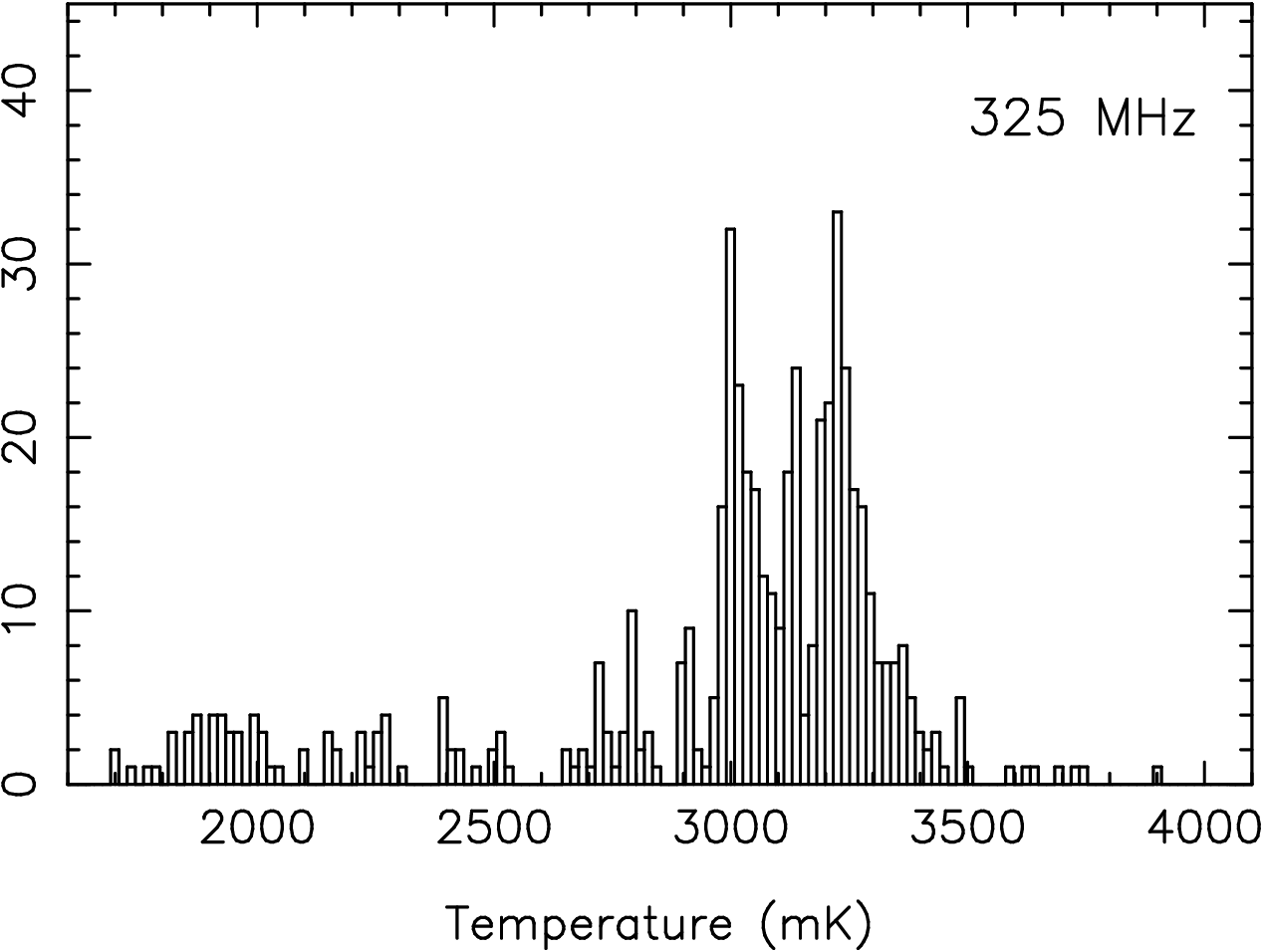}\includegraphics[width=55mm, height=65mm]{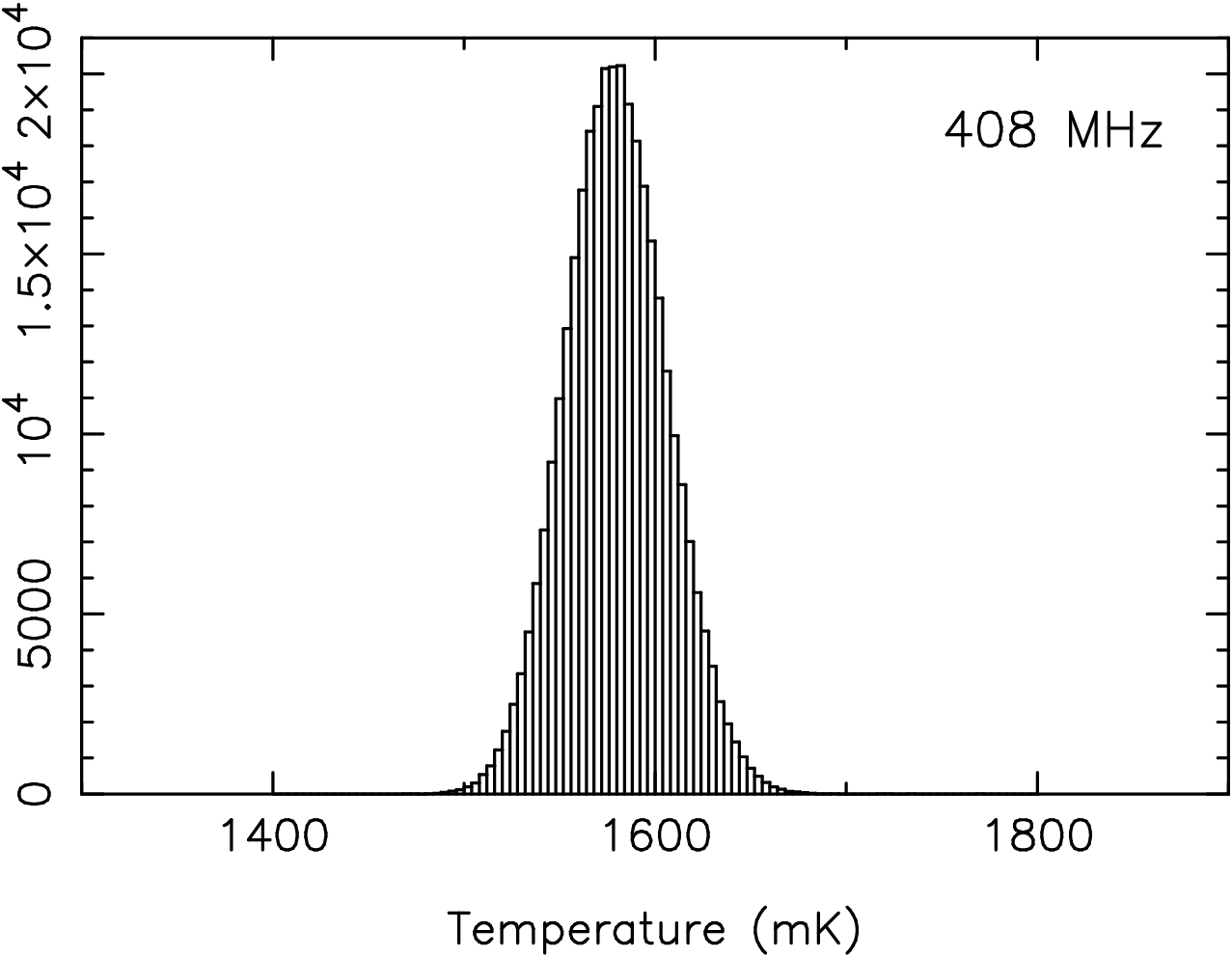}}
\rotatebox{270}{
\includegraphics[width=55mm, height=65mm]{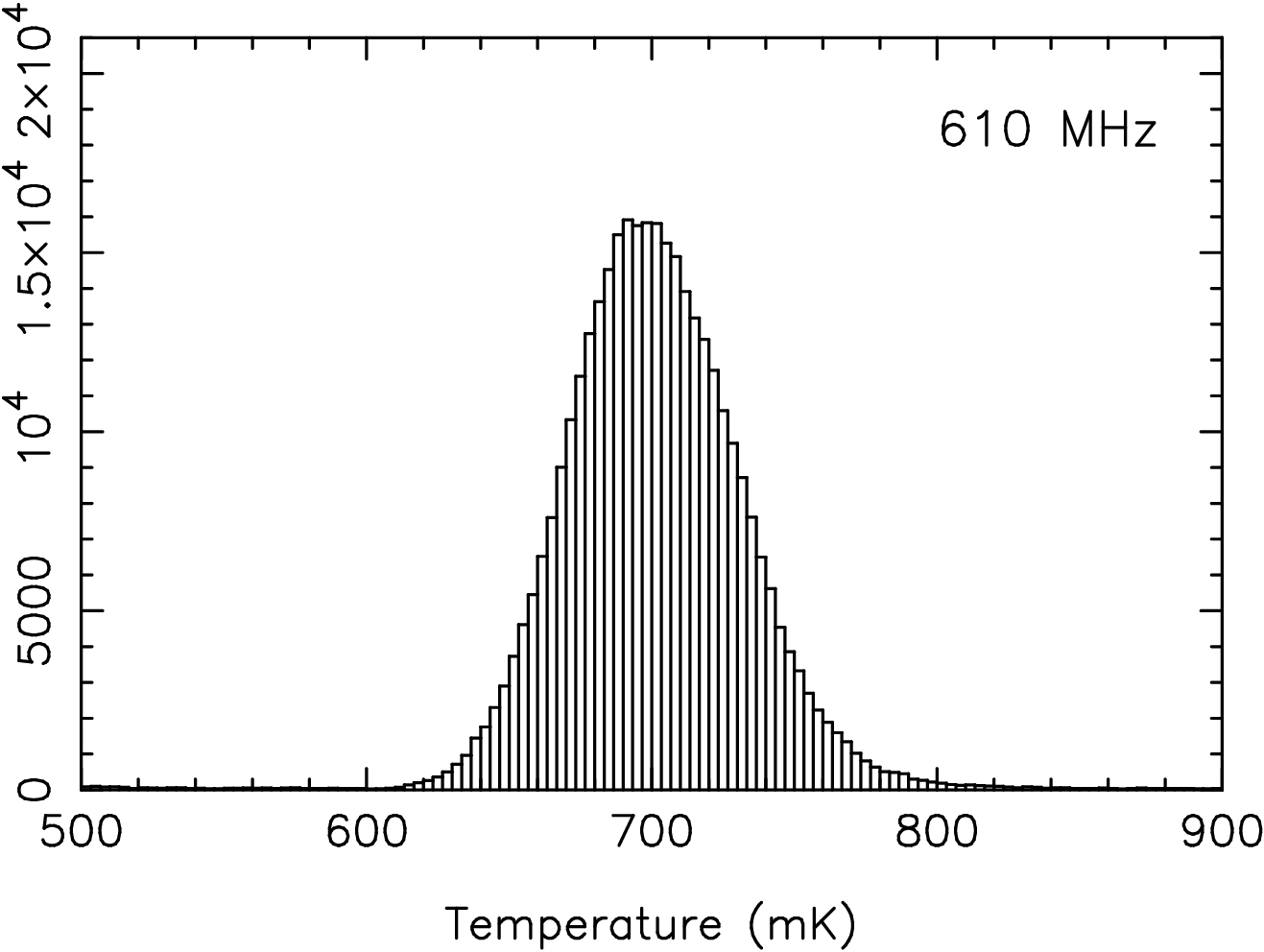}\includegraphics[width=55mm, height=65mm]{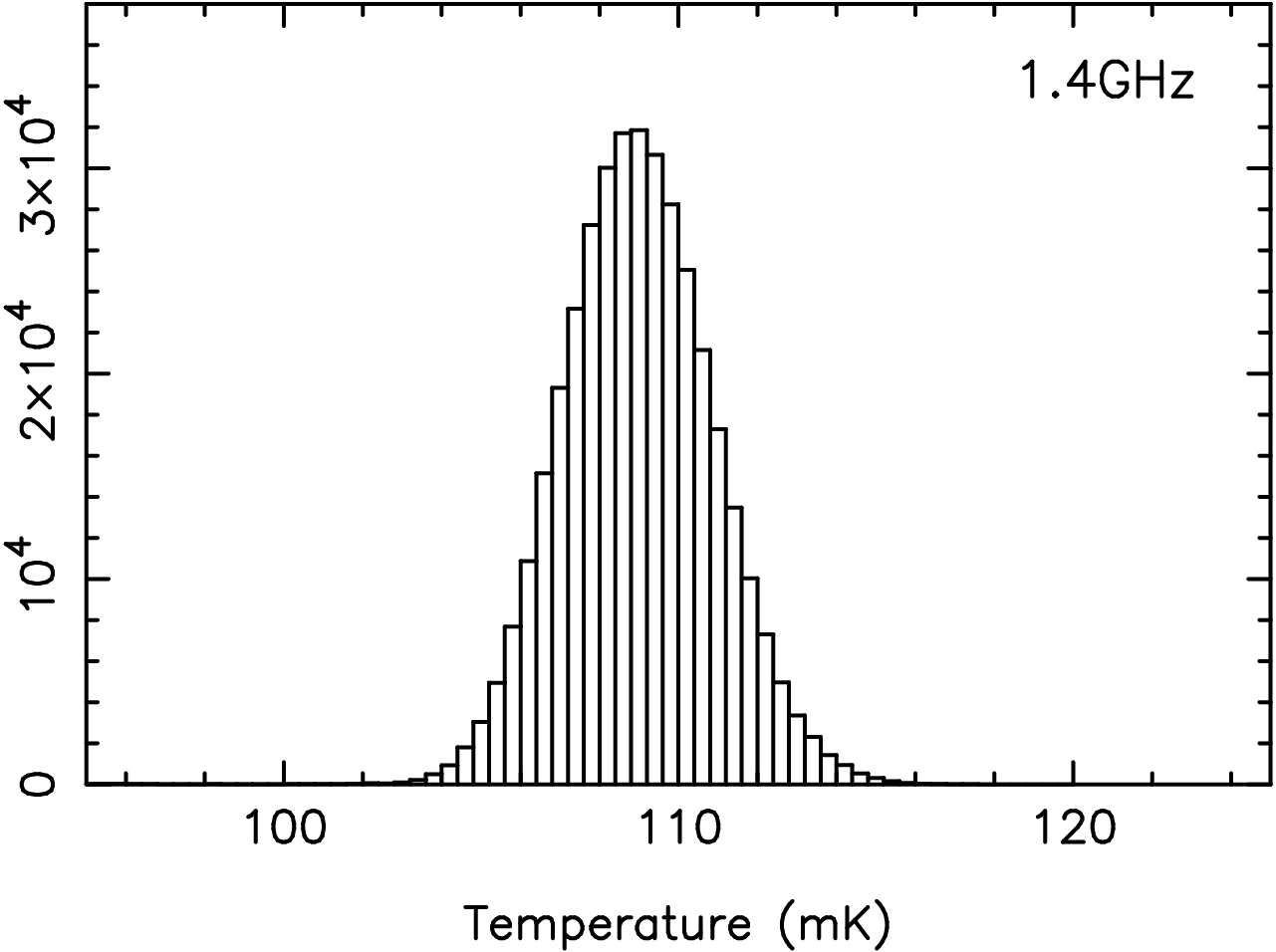}\includegraphics[width=55mm, height=65mm]{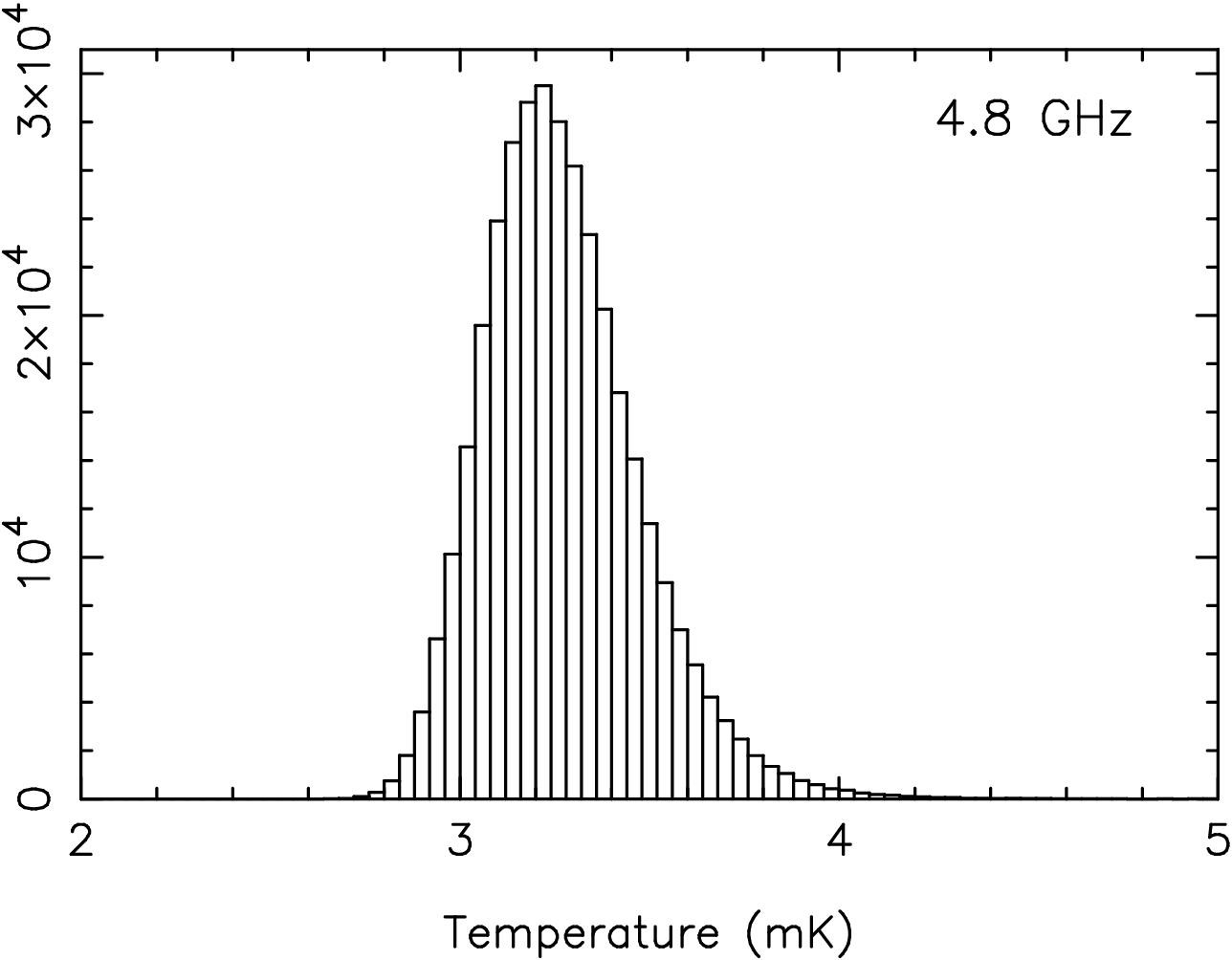}}
\rotatebox{270}{
\includegraphics[width=55mm, height=65mm]{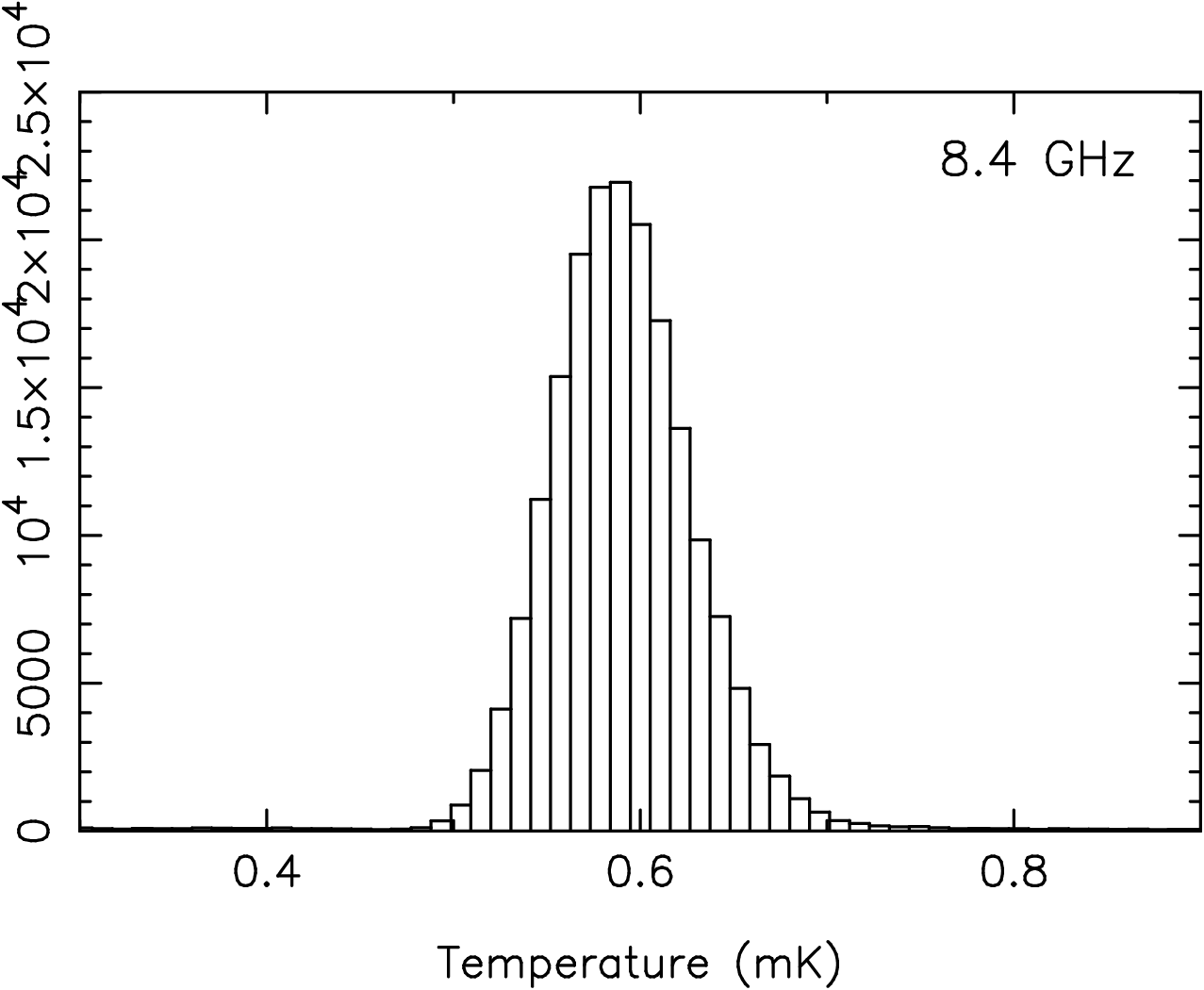}}
\caption{Histograms from the Markov chains at each frequency. 
         The background temperature was computed at each
        step in the chain and binned. At most frequencies there is a
        well defined value of the CRB with an approximately Gaussian
        distribution, while this is less true at $325\,$MHz in particular.}
\label{fig:markov}
\end{figure*}

Histograms of the chain values for the background temperature are
shown in Fig. \ref{fig:markov}. From the width
of these histograms we are able to measure the uncertainty in our 
estimates for the background temperature, taken here as 
the 68 percent area values, fully accounting for the correlations among
the parameters in the polynomial fits. The 1$\sigma$ uncertainties are listed in Table
\ref{tbl:temp}. 

Most of the histograms are fairly Gaussian, which
is a reflection of the quality of the data. Frequencies with good data
around the peak contribution (in the right-hand panels of Fig. 1) tend
to have well-constrained background temperature values, e.g. at $408\,$MHz. 
However, there is a noticeable irregularity with the $325\,$MHz histogram. 
Because of the limited data available at $325\,$MHz, and with the peak
area of contribution having little to no data, the 
histogram at this frequency does not have a well defined shape;
the uncertainty is far from Gaussian.

\subsection{Comparison with Previous Estimates}
\label{sec:previous}

Over the years there have not been many estimates of the CRB made using
source count
data (\citealp{Longair66,Pooley68,Wall90,Burigana04,Gervasi08}). And
even within this small list, the frequencies covered were rather
limited and uncertainties not always quoted. It is important to see
how our estimates compare with these previous
estimates. \citet{Longair66} gives a value for $T_{178}$ = 23 $\pm$ 5
K. \citet{Wall90} lists estimates of $T_{408} = 2.6\,$K, $T_{1.4} =
0.09\,$K, and $T_{2.5} = 0.02\,$ K. Our results are in agreement with
these earlier estimates to within $\pm$ 2$\sigma$. The values for source 
contributions from Gervasi et al. (2009) tend to be a
little higher than ours, the differences being traceable to choices
made for the limits of integration and for the parameterized form for
the fits.   

The ARCADE 2 experiment reported an excess of emission beyond what we and
others have estimated from source counts, with the excess largest at $3.4\,$GHz. 
We have also considered much lower frequencies in this paper than the $3.2\,$GHz
detection limit of ARCADE 2. However, it is possible to
calculate what temperatures would be expected using the best fit 
to the ARCADE 2 data:
\begin{equation}
T(\nu)=T_0 + A\left(\frac{\nu}{1\rm{GHz}}\right)^{\beta}\ .
\end{equation}
Here $T_0$ is the CMB base temperature, and the best fit values for the parameters
are $\beta$ = --2.56 and $A$ = 1.06 \citep{Seiffert09}. Measurements 
from the TRIS experiment were performed at $\nu$ = 0.6, 0.82, and
$2.5\,$GHz, and compared with the Gervasi et al. source contribution calculations are
within 3$\%$ at $0.6\,$GHz and 50$\%$ at $2.5\,$GHz

\begin{figure}

\includegraphics[scale=0.35,angle=270.0]{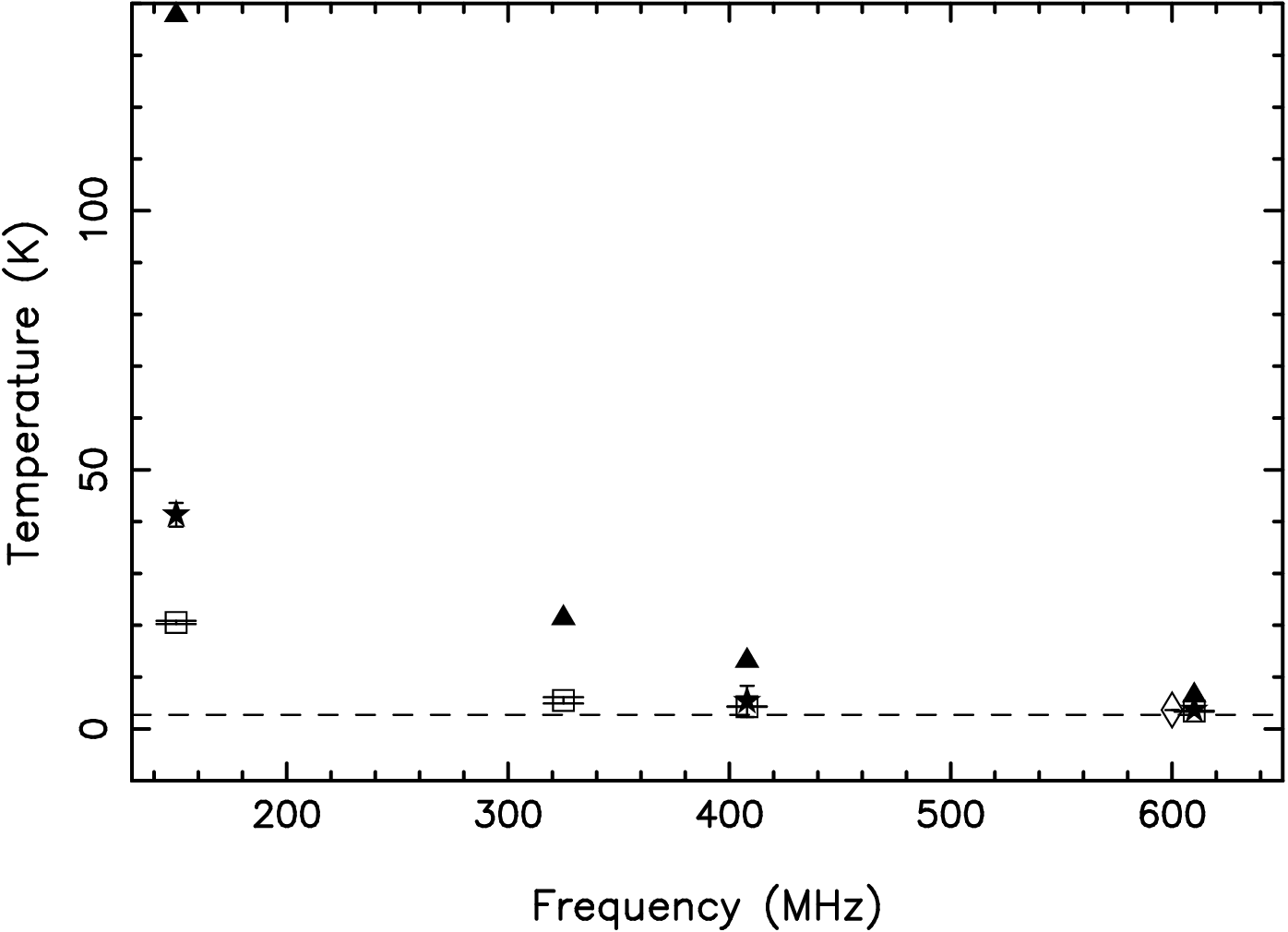}
\includegraphics[scale=0.35,angle=270.0]{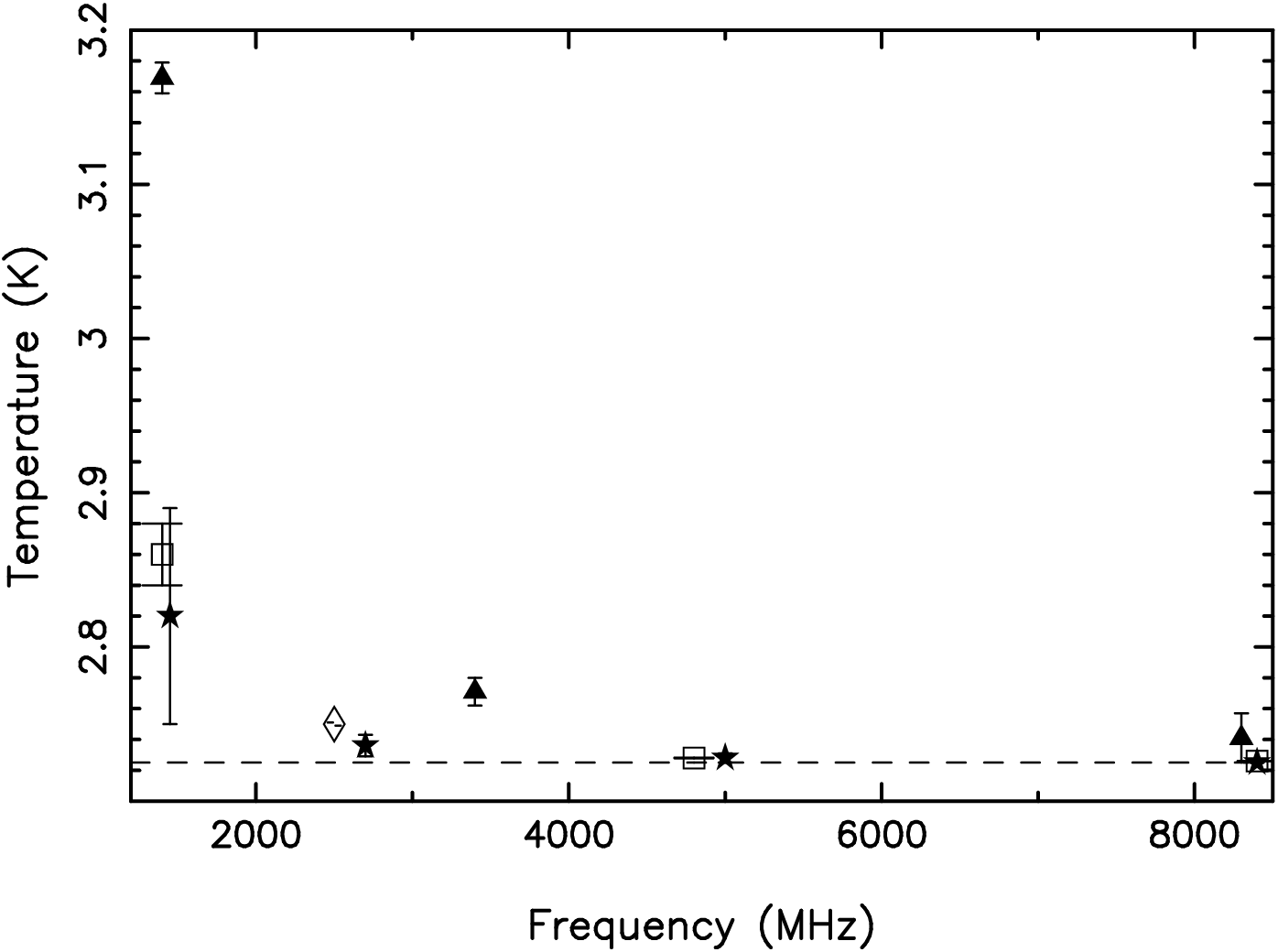}
\caption{Integrated sky brightness temperature at each
        frequency from the estimates in this paper (open squares), 
         Gervasi et al (2008, stars), TRIS measurements (Zannoni et
        al. 2008, diamonds), and ARCADE 2 measurements
        (Seiffert et al. 2009, triangles). The
        dashed line at the bottom represents the CMB temperature at
         $2.726\,$K \citep{Fixsen09a}.}
\label{fig:compare}
\end{figure}

The quantities detected by or extrapolated from ARCADE 2, those
estimated from counts by Gervasi et al. (2008), the measurements from the
TRIS experiment, as well as our current estimates are shown in
Fig.~\ref{fig:compare}. Here it can be seen that the ARCADE 2 absolute measurements 
lie far above both source estimates and TRIS measurements, particularly at
lower frequencies. Clearly, the excess detected around $3\,$GHz would  
correspond to a large excess at lower frequencies if the power-law continued.

\subsection{Systematic Errors}
\label{sec:errors}

We have considered several possibilities for systematic errors in exploring
whether our results might be compatible with those from the ARCADE 2
experiment. The first of these is possible bias from source clustering.
This can be an issue when dealing with surveys covering small areas, 
where one might get more field-to-field variations than expected from
Poisson errors. The two-point
angular correlation function for NVSS and FIRST
sources fits a power-law shape for separations up to at least
4$^\circ$ \citep{Blake02, Overzier03}. From this angular correlation 
one can estimate the fractional variance of the
counts \citep{Peebles80}. This procedure was
carried out by \citet{Dezotti09} and has been taken into account in the errors
provided and used in our estimations.

Another effect that could influence our results is the fact that in some
of the surveys used in our compilation the measurement frequency was
slightly different from the nominal one, i.e. 5 GHz rather than 4.8
GHz. In such cases we scaled the original measurements to the nominal
frequency using the assumed dependence of the source flux
$S(\nu) \sim \nu^{-0.7}$. This correction results in
negligible change in the derived fits.

An additional effect that could account for the difference in the background
temperatures is the possibility that some surveys have somehow missed
extended high-frequency emission blobs which could integrate up to the
required amounts. This seems an unlikely option, as such structures
would have to be on degree scales or larger to escape detection, and
because if these structures have features above a certain brightness
temperature then they would have been seen. 

Other possible effects to take into consideration for the
uncertainties include:

\begin{enumerate}
\item[1.] Calibration variations for different radio telescopes;
\item[2.] Inaccurate determination of completeness corrections at the
faint end;
\item[3.] Contribution from diffuse emission from the Intergalactic
Medium (IGM), Intercluster Medium (ICM), and the Warm-Hot
Intergalactic Medium (WHIM);
\item[4.] Missing low surface brightness emission from extended objects that
are either large, or sources with extended components, or sources that are
not detected if source surface brightness extends to low values. 
\end{enumerate}

\citet{Singal10} provide a detailed discussion of items 3 and 4 as well as
several other possibilities such as radio supernovae that could
contribute to the CRB.  We suspect that the most important effects are the first two items, particularly completeness at the faint end.

\section{Investigation of a Faint `Bump' in the Counts}
\label{sec:discussion}
At $1.4\,$GHz, where we have the most data, our estimated background
temperature plus a CMB baseline is $2.83\,$K $\pm$ $0.02\,$K, while an
extrapolation of the ARCADE 2 
result gives $3.17\,$K $\pm$ $0.01\,$K (error estimate from their
measurement at $3.2\,$GHz). This corresponds to a difference that is nearly
17$\sigma$ away from our estimate. It has been
suggested that this could be explained through an extra population of
faint radio galaxies, corresponding to a `bump' in the
Euclidean-normalized counts at flux densities near or below where the
current data are petering out (see also \citealt{Singal10}). We want to investigate how big
this bump would need to be in order to explain the excess emission.

\begin{figure}
\includegraphics[scale=0.35, angle=270.0]{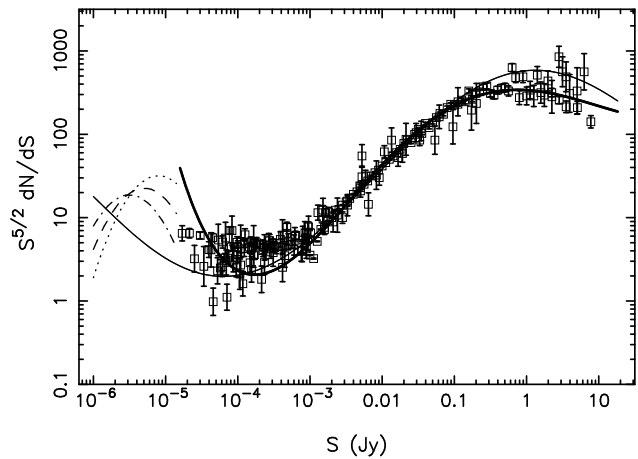}
\caption{The 1.4 GHz source count data. The thin solid line gives the best
        fit to the data while having a moderate slope at the faint end
        and integrating to the necessary excess reported by ARCADE 2.
        The thick solid line is our best fit to the 1.4 GHz data from
        Section 2. The other three lines are bumps
        peaking at 7.9 (dotted), 5.0 (dashed), and 3.1 (dot dash)
        $\mu$Jy, which produce the background temperature
         necessary to match the ARCADE 2 results. On this plot the
        height of such a bump is proportional to $S^{1/2}_{\rm{peak}}$}
\label{fig:bump}
\end{figure}

We carried out two separate approaches for modelling such a bump, 
the outcomes of which can be seen in Fig. \ref{fig:bump}.
Our first approach is a simple extension of the current counts with an upward
trend below 10 $\mu$Jy, but one not quite as steep as the best fit
line. To do this we simply added artificial data points past the lower
flux density limit of the rest of the data in order to control the end 
behaviour of the fit line. We then investigated what was required
to match the ARCADE 2 results. The solid line of
Fig. \ref{fig:bump} shows the results of this fitting. It is the best
fit to the data, allowing for a moderate upward slope in the faint
end. When integrated from $10^{-6}$Jy the result is enough to account for
the temperature reported by ARCADE 2. 

Our second method involved choosing a simple parabola with fixed width
of a decade in log-$S$ and variable position for the peak,
and running a Markov chain that
fit the height parameter that would 
integrate to give the amount of excess emission needed to match the 
ARCADE 2 result. We found that the peak of the
bump could be at flux densities as high as 8.0 $\mu$Jy.

It is relatively easy to produce a bump
big enough to account for the extra emission while still fitting 
the rest of the data reasonably well, with either method.
However, we do know that any such bump is
constrained by the observed IR background, through the IR-Radio
correlation (see e.g. \citealt{Haarsma98}). This correlation will 
have to be taken into account in any modelling of this faint flux 
density bump so as not to overproduce the IR background. This essentially
requires any faint radio population to be quite IR faint compared with
known galaxy types.

\section{Conclusions}

We used source count data from $\nu$ = $150\,$MHz, $325\,$MHz, $408\,$MHz, 
$610\,$MHz, $1.4\,$GHz, $4.8\,$GHz, and $8.4\,$GHz to evaluate the
contribution from sources to the diffuse cosmic radio background. Polynomials
were fit to the data and integrated to obtain lower bound estimates at each
frequency for the sky brightness temperature. In addition, we also
extrapolated our fits beyond the limits where data are available using
reasonable assumptions for how the curves behave in
those regions. We then used Monte Carlo
Markov Chains to obtain estimates of the uncertainties of the
temperature estimates at each frequency and also considered other
possible sources of uncertainties that could affect the results. 

Our estimates are considerably lower than the
measurements of ARCADE 2, even when taking into account the
uncertainties or extrapolations. We considered the possibility
that the excess emission comes from a
bump in the source counts in the $\mu$Jy range at $1.4\,$GHz. We used modelling to
see how large such a bump must be in order to obtain the necessary
contribution to the background. We saw that a bump could exist in this
range, peaking at fluxes as bright as 8 $\mu$Jy, and could integrate up to the
excess emission of $\pm$~$320\,$mK, with a height that is consistent with
the data. 

We still have no direct evidence that such a new population exists, and
so further investigation into the faint end of the counts is needed.
The infrared and radio connection
could be used to test this idea through use of signal stacking and by
examining different possible luminosity functions to look at the
evolution of such a population. The final answer may only
be reached when source count data become available in the $\mu$Jy
range,  perhaps in the era of the EVLA and eventually 
the SKA.

\section*{Acknowledgments}
This research was supported by the Natural Sciences and Engineering
Research Council of Canada.

\label{lastpage}
\end{document}